\documentclass[preprint]{aastex}

\usepackage{epsfig}

\received{ }
\accepted{ }
\shorttitle{A Rapid Flare From Markarian 501}
\shortauthors{Catanese and Sambruna}

\begin{document}

\title{A Rapid X-ray Flare from Markarian 501}

\author{Michael Catanese}
\affil{Fred Lawrence Whipple Observatory,
Harvard-Smithsonian CfA, P.O. Box 97, Amado, AZ 85645-0097}
\email{mcatanese@cfa.harvard.edu} 

\and 

\author{Rita M. Sambruna}
\affil{Pennsylvania State University, Department of
Astronomy and Astrophysics, 525 Davey Lab, State College, PA 16802}

\begin{abstract} 

We present X-ray observations of the BL Lacertae (BL Lac) object
Markarian 501 (Mrk\,501), taken with the {\it Rossi X-ray Timing
Explorer} in 1998 May as part of a multi-wavelength campaign.  The
X-ray light curve shows a very rapid flare in which the 2-10 keV flux
increased by $\sim$60\% in $<$200 seconds.  This rapid rise is
followed by a drop-off in the 2-10 keV flux of $\sim$40\% in $<$600
seconds.  The 10-15 keV variation in this flare is roughly a factor of
two on similar time-scales.  During the rise of the flare, the 3-15
keV spectral index hardened from 2.02$\pm$0.03 to 1.87$\pm$0.04, where
it remained during the decay of the flare.  This is the fastest
variation ever seen in X-rays from Mrk\,501 and among the fastest seen
at any wavelength for this object.  The shift in the energy at which
the spectral power peaks (from $\lesssim$3 keV before the flare to $/gtrsim$30
keV during the flare) is also among the most rapid shifts seen from
this object. This flare occurs during an emission state (2-10 keV flux
$\approx 1.2 \times 10^{-10}$ erg cm$^{-2}$ s$^{-1}$) that is
approximately 25\% of the peak flux observed in 1997 April from this
object but which is still high compared to its historical average
X-ray flux.  The variations in the hardness ratio are consistent with
the low energy variations leading those at high energies during the
development and decay of the flare.  This pattern is rare among high
frequency peaked BL Lac objects like Mrk 501, but has been seen
recently in two other TeV emitting BL Lacs, Mrk 421 and PKS 2155-304.
The hard lag is consistent with a flare dominated by the acceleration
time-scale for a simple relativistic shock model of flaring.

\end{abstract}


\section{Introduction}
\label{intro}

BL Lacertae objects (BL Lacs) are members of the blazar class of
active galactic nuclei (AGN).  Blazars exhibit rapid, large amplitude
variability at all wavelengths, high optical and radio polarization,
apparent superluminal motion, and in some cases, gamma-ray emission.
All of these observational properties lead to the broadly held belief
that blazars are AGN with jets oriented nearly along our line of
sight. The broadband spectral energy distribution of blazars, when
plotted as $\nu$F$_\nu$ versus frequency, shows a double peaked shape,
with a smooth extension from radio to between IR and X-ray frequencies
(depending on the specific blazar type), followed by a distribution
that typically starts in the X-ray band and can peak in the gamma-ray
band, at energies as high as several hundred GeV.  The low energy part
is believed to be incoherent synchrotron radiation from a relativistic
electron-positron plasma in the blazar jet.  The origin of the high
energy emission is still a matter of considerable debate (e.g.,
Buckley 1998; Mannheim 1998).

Markarian 501 (Mrk 501) is one of the closest BL Lacs ($z=0.034$)
known and it is one of the brightest in the X-ray band.  Because its
peak spectral power output occurs at UV/X-ray energies, Mrk 501 is
classified as a high frequency peaked BL Lac.  It is one of only six
BL Lacs reported as sources of very high energy (VHE, E $>$ 250 GeV)
gamma rays and, along with Mrk 421, one of only two that have been
confirmed as VHE sources (for a review, see Catanese \& Weekes
1999). Mrk 501 has also been detected as a source of gamma rays at GeV
energies with the Energy Gamma-Ray Experiment Telescope (EGRET;
Kataoka et al. 1999) and at a few hundred keV by the Oriented
Scintillation Spectrometer Experiment (OSSE; Catanese et al. 1997) on
the {\it Compton Gamma-Ray Observatory}.

Just as all other blazars, Mrk 501 exhibits rapid, large amplitude
variability over a wide range of wavelengths.  In X-rays, variations
of from 30\% to 300\% were observed on time scales of days in 1997
during a high emission state \citep{Pian98,Lamer98,Catanese99b} but no
sub-day scale flares were seen.  The fastest variation observed in
X-rays was a flux increase of approximately 20\% in about 12 hours
seen with EXOSAT in 1986 \citep{Giommi90}.  Spectral variability in
X-rays from Mrk 501 has been both moderate, with changes in the
spectral index of $\sim$0.1-0.3 on several day scales (e.g., Pian et
al. 1998), and rapid, with spectral variations of $\sim$0.5 on 2-3 day
time-scales, observed in 1998 June \citep{Sambruna99a}.  The Whipple
Observatory has observed VHE gamma-ray variations spanning a factor of
$>$70 in flux in four years of observations and has observed
significant variability with time-scales as short as 2 hours
\citep{Quinn99}.  Similar variability ranges are observed by other VHE
telescopes \citep{Hayashida98,Aharonian99a,Djannati99}.  In the
R-band, \citet{Miller99} reported the detection of a flare in which
the flux increased by 4\% (from $V_R = 13.90$ to 13.80) in 15 minutes
with a decay to the previous level in the same amount of time.

Multi-wavelength observations have revealed correlations between VHE
gamma rays and X-rays in this object \citep{Catanese97} and in 1997,
the synchrotron spectrum of Mrk 501 was observed to extend up to
approximately 100 keV \citep{Catanese97,Pian98,Catanese99b}, the
highest seen in any blazar and a 50-fold increase over what was
observed only one year before \citep{Kataoka99}.  This behavior has
established Mrk 501 as the prototype for a subset of BL Lacs that
exhibit large shifts in the peak of their synchrotron spectra during
flares.

In this paper, we report on observations of Mrk 501 taken with the
{\it Rossi X-ray Timing Explorer} ({\it RXTE}) in 1998 May as part
of a multi-wavelength campaign.  The full multi-wavelength results
will be reported in a future work.  Here, we concentrate on the
observation of a very rapid flare and discuss its implications.

\section{Observations and Analysis}
\label{observe}

{\it RXTE} consists of the Proportional Counter Array (PCA; Jahoda et
al. 1996) which is sensitive to 2-60 keV photons, the High Energy
X-ray Transient Experiment (HEXTE; Rothschild et al. 1998) which is
sensitive to 15-150 keV photons, and the All-Sky Monitor
(ASM; Levine et al. 1996) which is sensitive to 2-12 keV photons.  Here
we report on the results of the PCA observations.  The count rate for
HEXTE was too low to obtain significant count rate variations on the
time scales observed.  There are no observations with the ASM during
the period of the flare.  Also, because the X-ray flare occurred at
approximately 18 hours Universal Coordinated Time (UTC), there are no
overlapping TeV observations from Whipple, HEGRA, CAT, or the
Telescope Array.

The observations of Mrk 501 occurred between 1998 May 15 and 29.  They
consisted of roughly four 2-3 ks pointings per day for the full two
weeks.  After screening for good time intervals, as described below,
the data set consists of 114 ks of observations.  Each observation
resulted in a very significant detection of Mrk 501 that allowed
spectra to be resolved and short-term variability to be investigated.

We used FTOOLS v4.2 to analyze this data.  The background was
estimated using the weak source background models (appropriate for
sources with count rates $<$40 counts/s/PCU) and the latest response
matrices obtained from the {\it RXTE} Guest Observer Facility (GOF)
web
site.\footnote{http://heasarc.gsfc.nasa.gov/docs/xte/xte$_-$1st.html}
Good time intervals were selected from the standard 2 data files using
the screening criteria recommended by the {\it RXTE} GOF.  Finally,
only proportional counter units (PCUs) 0, 1, and 2 were active for the
vast majority of the observations presented here, so we only use those
PCUs in this work.  For spectra and all other light curves, we used
all three layers.  Spectral fits were performed using XSPEC 10.0.  We
use the Galactic hydrogen column density of $1.73 \times 10^{20}$
cm$^{-2}$ \citep{Elvis89} to model the effect of photoelectric
absorption, which is negligible for the energy range covered by the
PCA for this object located far from the Galactic plane.

\section{Lightcurves}
\label{lc}

\begin{figure*}[t]
\centerline{\epsfig{file=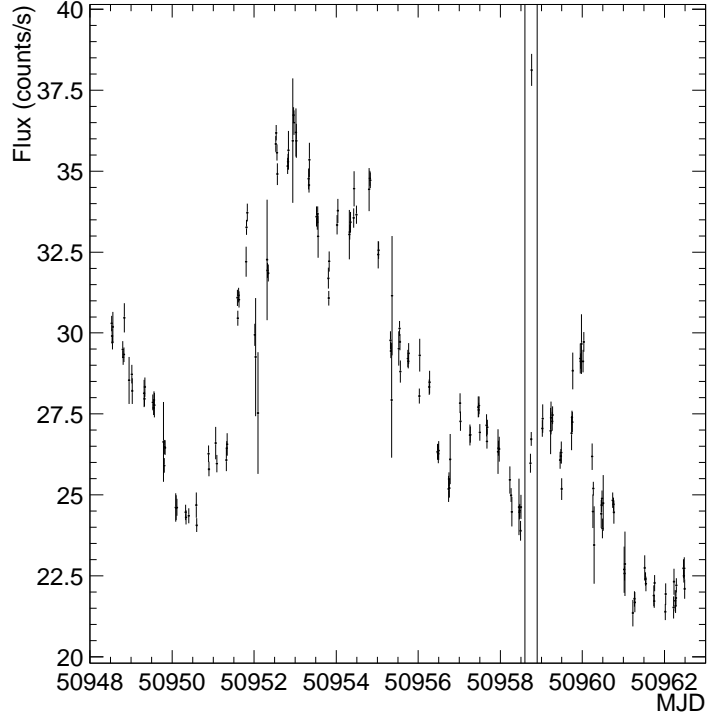,height=3.8in}}
\caption{The 2-10 keV lightcurve for Mrk 501 between 1998 May 15 - 29
as measured with the {\it RXTE} PCA.  The flux is shown in 1024-second
bins in units of counts per second and the horizontal axis is the
Modified Julian Day for the observations.  The two vertical lines
indicate the observation region where there is a very rapid flare.
\label{1998_lc}
}
\end{figure*}

The 2-10 keV lightcurve for the 1998 May observations is shown in
Figure~\ref{1998_lc}.  The data points are shown in 1024-second bins.
The count rate varies by approximately a factor of 1.8 from 21 cts/s
to 38 cts/s.  By comparison, {\it RXTE} observations of Mrk 501 in
1997 April exhibited 2-10 keV lightcurve count rates of 80 cts/s to
160 cts/s \citep{Catanese99b}.  Thus, Mrk 501 was in a much lower
emission state during these observations than in 1997, though the flux
was relatively high by historical standards.  The light curve begins
with a 20\% drop over two days, follows with a 50\% rise over three
days, and ends with a gradual decline in flux of 65\% over the
remaining nine days of the observation.  During the decline, small
amplitude, day-scale flares are evident at MJD 50955 and 50960.

Most notably, toward the end of MJD 50958 (indicated by the two
vertical lines in Fig.~\ref{1998_lc}), the flux is much higher than
the surrounding observations, and is actually the highest count rate
in this light curve.  For the remainder of the paper we concentrate on
this flare observation.  Detailed analysis of the entire 1998 data set
will be presented in a future paper.  This flare occurs during what is
otherwise a somewhat unremarkable observation period.  There is no
evidence of significant short term variability within any other single
observation nor is the variability seen on day scales in the other
observations of such large amplitude.  The largest variation on any
other day is a 30\% increase of the flux between MJD 50951 and 50952.
Our comparisons of background files with the PCA background model and
investigations of the data quality monitors (e.g., electron
activation) confirm that this flare is not a spurious effect.

\begin{figure*}
\centerline{\epsfig{file=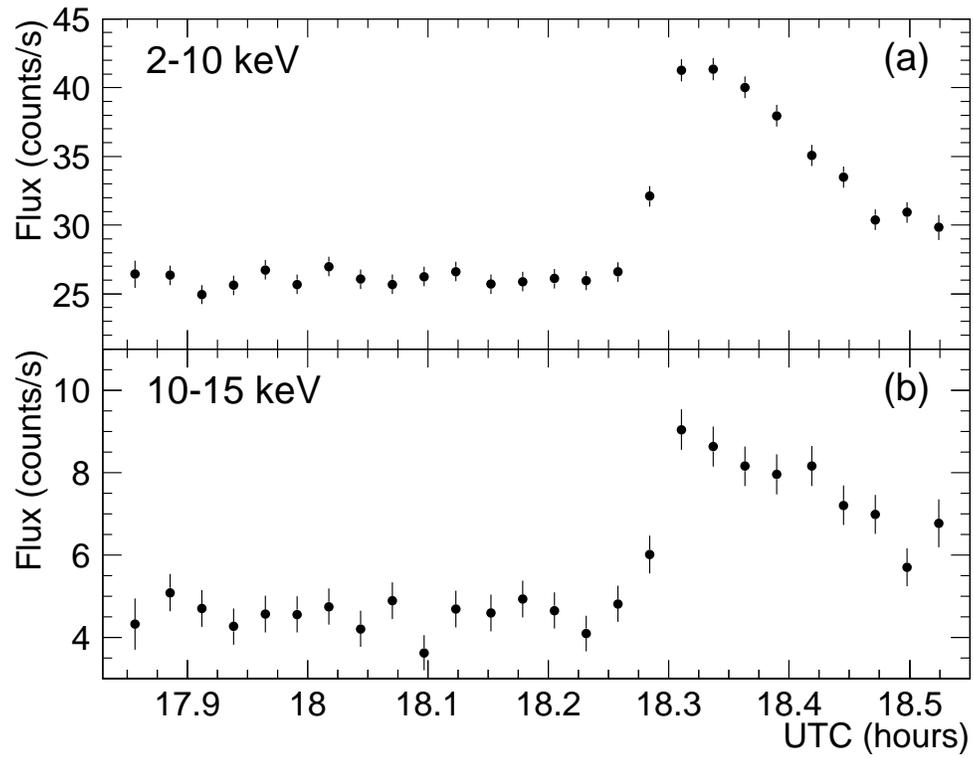,height=4.0in}}
\caption{The 2-10 keV (a) and 10-15 keV (b) light curves of Mrk 501
between 17.8 and 18.6 hours of 1998 May 25 as measured with the
{\it RXTE} PCA.  The flux is shown in 96-second bins in units of
counts per second and the horizontal axis is the hours of the day in
Universal Coordinated Time (UTC).
\label{flare_lc}
}
\end{figure*}

In Figure~\ref{flare_lc}, we show the 2-10 keV and 10-15 keV light
curves for the observation indicated by the vertical lines in
Figure~\ref{1998_lc}.  The observation was taken on May 25 between
17.9 and 18.6 hours Universal Coordinated Time (UTC).  The data are
shown in 96-second bins.  During this observation, Mrk 501 exhibited
low flux, very steady emission for about half of the observation.
After this, the count rate increased from approximately 26 cts/s to 41
cts/s in approximately 200 seconds, corresponding to a brightness
increase of 13\%/minute.  This is followed by a steady decrease in the
count rate to approximately 30 cts/s in approximately 580 seconds, a
decline rate of 3\%/minute.  Though not quite as well measured due to
lower statistics, the flare is clearly evident in the 10-15 keV light
curve, with a similar rise and fall time-scale.  The 10-15 keV flux
variation rate is 18\%/minute and 3\%/minute for the rise and fall of
the flare, respectively.

Observations were also taken with {\it RXTE} approximately 5.5 hours
before and 5.5 hours after this observation. Neither showed any
significant variability.  The observation before the flare had an
average 2-10 keV count rate of 24.5 cts/s and the one after the flare
had an average count rate of 27 cts/s.  Both are consistent with
little or no change in flux compared to that seen at the start and end
of the flare observation.

\section{Spectra}
\label{spectra}

The average spectrum for the observations between May 15 and 29 is
best fit by a broken power law model with a photon spectral index of
1.92$\pm$0.01 up to a break at 6.2$\pm$0.3 keV, above which the
spectral index is 2.07$\pm$0.01.  The 2-10 keV flux during this period
is $(1.12 \pm 0.02) \times 10^{-10}$ erg cm$^{-2}$ s$^{-1}$.  The
spectrum extends to at least 40 keV with no evidence of another break.
This spectrum connects smoothly with OSSE measurements during this
period \citep{Buckley99}.  Observations taken with {\it Beppo SAX} on
April 28 and 29 and May 1 \citep{Pian99} indicate a somewhat harder
spectrum with peak power output at $\sim$20 keV.  The X-ray flux
during the {\it Beppo SAX} observation is $\gtrsim$50\% higher than in
the {\it RXTE} observations reported here, so the spectral shift is
consistent with the previously observed tendency of Mrk 501 to
increase the energy at which the spectral energy distribution peaks as
its flux increases (e.g., Pian et al. 1998).

Spectral analysis of the flare observation reveals a rapid change
during the course of the flare.  We fit the spectrum from 3-15 keV
where there are sufficient photon statistics and no problems with the
PCA response function.  The average spectrum for the observation is
well-fit by a simple power law with a photon spectral index of $\Gamma
= 1.95 \pm 0.03$.  To investigate the evolution of the spectrum during
the course of the flare, we break up the data into three regions:
before the flare, during the rise, and during the decay.  All are
well-fit by simple power laws and a summary of those fits is given in
Table~\ref{spec_fits}. The photon spectral index is $\Gamma = 2.02 \pm
0.03$ before the flare, $2.04 \pm 0.11$ during the rise of the flare,
and $1.87 \pm 0.04$ during the decay of the flare.  The spectrum
during the flare follows this power law at least out to 30 keV,
indicating a large shift in the location of the peak power output
during the flare.  Dividing the decay of the flare into two parts does
not reveal significantly different spectra than the average spectrum
for the entire decay region.

The observation taken approximately 5.5 hours before the flare
indicates a spectral index of $2.02 \pm 0.02$ and a 2-10 keV flux of
$(0.98 \pm 0.04)\times 10^{-10}$ erg cm$^{-2}$ s$^{-1}$, consistent
with the observations just prior to the flare.  The observation taken
approximately 5.5 hours after the flare observation indicates a
spectral index of $2.08 \pm 0.04$ and a 2-10 keV flux of $(1.10 \pm
0.08) \times 10^{-10}$ erg cm$^{-2}$ s$^{-1}$, indicating a
significant softening of the spectrum following the flare.

\section{Discussion}
\label{discuss}

The variation in the spectral index during the course of the flare can
provide insights into the dominant flaring timescales and acceleration
process.  As discussed by \citet{Kirk99}, for a flare in which the
variability and acceleration time-scales are much less than the
cooling time-scale a plot of the spectral index versus flux should
follow a clockwise pattern, i.e., the harder energies vary first.  For
a flare where the variability, acceleration, and cooling time-scales
are similar, the spectral index versus flux diagram should move in a
counter-clockwise direction, i.e., the softer energies vary first
because the number of particles changes due to the acceleration
process which proceeds from low energy to high energy.  Clockwise
patterns are most commonly observed in the TeV sources Mrk 421 (e.g.,
Takahashi et al. 1996) and PKS 2155-304 (e.g., Kataoka et al. 2000)
but counter-clockwise patterns have been recently observed from these
objects \citep{Fossati99,Sambruna99b}.

\begin{figure*}
\centerline{\epsfig{file=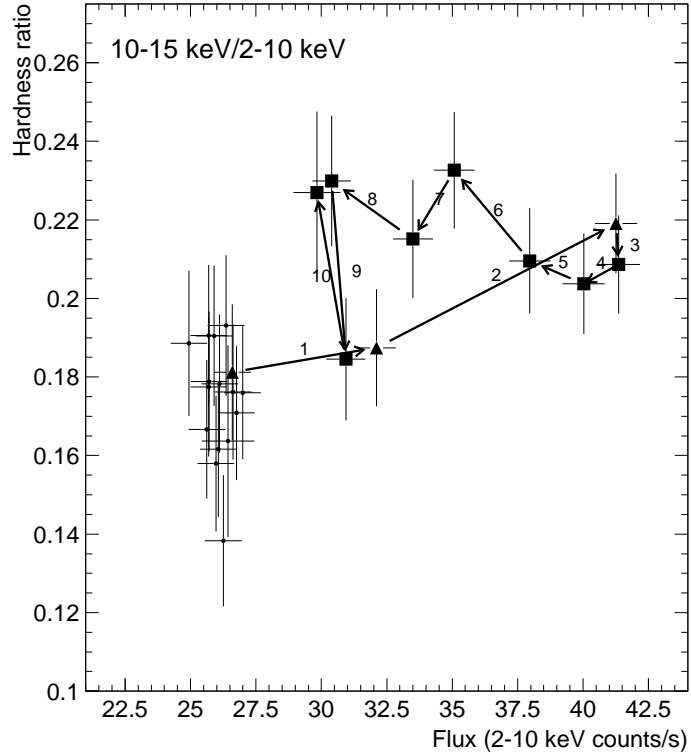,height=4.0in}}
\figcaption{The hardness ratio (10-15 keV count rate/2-10 keV count rate)
versus 2-10 keV count rate for observations of Mrk 501 between 17.8
and 18.6 hours (UTC) of 1998 May 25 as measured with the {\it RXTE}
PCA.  Each point represents one 96-second bin.  The filled circles
indicate data from before the onset of the flare.  The numbers follow
the progression of the hardness ratio in time during the flare.  
Thus, the arrow marked with a ``1'' shows the
onset of the flare and the arrow marked with a ``10'' indicates the
last time bin of the observation.  The filled triangles indicate data
taken during the rise of the flare and the filled squares indicate
data taken during the decay of the flare.
\label{hard_flux}
}
\end{figure*}

Because the data do not have sufficient statistics to plot spectral
index versus flux on such short time scales, we instead plot the
hardness ratio (10-15 keV count rate/2-10 keV count rate) versus flux
for the flare on May 25 in Figure~\ref{hard_flux}.  The numbers in the
plot indicate the development of the hardness ratio in time during the
flare.  The large cluster of filled circles represents the
observations before the onset of the flare.  The filled triangles
represent the rising part of the flare.  The point indicated by the
tail of the arrow marked with the ``1'' is the last low flux point
before the flare starts.  The filled squares are data taken during the
decay of the flare.  During the rise of the flare, the hardness ratio
increases steadily.  During the decay of the flare, there is a slight
trend for the hardness ratio to increase even further.  Thus, these
observations are consistent with a counter-clockwise pattern.  A
clockwise pattern seems precluded by the significantly harder spectrum
during the decay of the flare than the rise (see
Table~\ref{spec_fits}) but other patterns in the hardness ratio versus
flux diagram cannot be ruled out given the statistical errors in this
observation.  The counter-clockwise pattern and the large shift in the
synchrotron peak imply that the acceleration process dominates the
development of the flare, accelerating a fresh population of high
energy electrons which causes the flare.

In summary, {\it RXTE} observations of Mrk 501 in 1998 May have
revealed a flux state which was approximately one-fourth as strong as
observed in 1997, and an average spectrum with peak power output at
approximately 6 keV.  This is a decrease of more than a factor of 15
from the 100 keV peak seen in 1997.  During these observations, a very
rapid flare was observed in which the location of peak power output
increased from $\lesssim$3 keV to $gtrsim$30 keV.  This large shift in
peak power output energy is similar to the behavior of Mrk 501 in 1997
and in 1998 June.  The evolution of the hardness ratio of the flare is
consistent with the flare development being dominated by the
acceleration process but the lack of simultaneous multi-wavelength
observations prohibits further detailed testing of emission models.

Though rapid variations have been seen from Mrk 501 at other
wavelengths before, it was generally regarded as a more slowly varying
object than the other TeV sources, Mrk 421 and PKS 2155-304.  Thus,
one could conduct less dense observations of Mrk 501 and still sample
the variations with adequate coverage to resolve the shape of the
variations and the correlations between wavelengths.  These
observations show that very dense multi-wavelength observations are
required for Mrk 501 as well since it can vary on time-scales
comparable to the fastest seen in the other TeV sources.  They also
indicate that, as has been seen in Mrk 421 \citep{Gaidos96}, these
very rapid flares can occur when the source is not in a particularly
high emission state, so dense observations must be used regardless of
the flux level observed from these objects.  Multi-wavelength
observations of such rapid flares will provide stringent tests of
emission models for these TeV sources and may lead to a better
understanding of the acceleration process that occurs in their jets.

\acknowledgments 

The authors wish to thank K. Jahoda and T. Jaffe for advice about the
{\it RXTE} analysis.  MC wishes to thank T. Weekes, D. Carter-Lewis,
J. Finley, J. Buckley, C. Dermer, N. Johnson, and F. Krennrich for
their support of these observations.  MC acknowledges grant support
from NASA and the U. S. Department of Energy.  RMS acknowledges 
support from NASA contract NAS--38252.

%\clearpage

%\clearpage

\begin{deluxetable}{cccc}
\tablewidth{0pt}
\tablecaption{Spectral fits of 1998 May 25 flare data \label{spec_fits}}
\tablehead{\colhead{UT(hrs)} & \colhead{$\Gamma$} & \colhead{Flux 
\tablenotemark{a}} & \colhead{$\chi^2$/(d.o.f.)}}
\startdata
17.858 -- 18.524 & 1.95$\pm$0.03 & 11.90$\pm$0.09 & 0.895 (30) \\
17.858 -- 18.284 & 2.02$\pm$0.03 & 10.46$\pm$0.11 & 0.768 (30) \\
18.284 -- 18.311 & 2.04$\pm$0.11 & 13.34$\pm$0.46 & 0.483 (30) \\
18.311 -- 18.524 & 1.87$\pm$0.04 & 14.50$\pm$0.15 & 0.571 (28)\tablenotemark{b} \\
\enddata
\tablenotetext{a}{2-10 keV flux listed in units of 10$^{-11}$ erg cm$^{-2}$ 
s$^{-1}$.}
\tablenotetext{b}{An absorption edge was included in this fit because
activation of the Xenon gas 
was not completely eliminated by the response
function and background model, producing an absorption edge at 4.8 keV.}
\end{deluxetable}

\newpage

\end{document}